\documentclass[authoryear,preprint,12pt]{elsarticle}
\usepackage{amsfonts}

\usepackage{amssymb}

\journal{Ecological Modelling}
\begin{document}

\begin{frontmatter}



\title{Common Patterns of Energy Flow and Biomass Distribution on Weighted Food Webs}


\author{Jiang Zhang, Yuanjing Feng}

\address{Department of Systems Science, School of Management, Beijing Normal University, Beijing 100875, China}

\begin{abstract}
Weights of edges and nodes on food webs which are available from the
empirical data hide much information about energy flows and biomass
distributions in ecosystem. We define a set of variables related to
weights for each species $i$, including the throughflow $T_i$, the
total biomass $X_i$, and the dissipated flow $D_i$ (output to the
environment) to uncover the following common patterns in 19
empirical weighted food webs: (1) DGBD distributions (Discrete
version of a Generalized Beta Distribution), a kind of deformed
Zipf's law, of energy flow and storage biomass; (2) The allometric
scaling law $T_i\propto X_i^{\alpha}$, which can be viewed as the
counterpart of the Kleiber's $3/4$ law at the population level; (3)
The dissipation law $D_i\propto T_i^{\beta}$; and (4) The gravity
law, including univariate version $f_{ij}\propto (T_iT_j)^{\gamma}$
and bivariate approvement $f_{ij}\propto
T_i^{\gamma_1}T_j^{\gamma_2}$. These patterns are very common and
significant in all collected webs, as a result, some remarkable
regularities are hidden in weights.
\end{abstract}


\end{frontmatter}


\section{Introduction}
\label{sec.introduction}

Complex network is a useful tool to study interactions and
relationships between components of a complex
system\citep{watts_collective_1998,albert_statistical_2002}. In
ecology, complex network models are used in many ways, including
representing trophic interactions in food
webs\citep{montoya_small_2002,dunne_food-web_2002,williams_two_2002,berlow_interaction_2004,emmerson_predator-prey_2004}
and energy-matter flux in
ecosystems\citep{odum_self-organization_1988,finn_measures_1976,szyrmer_total_1987,higashi_extended_1986,higashi_network_1993,baird_seasonal_1989,fath_review_1999}.
Some common patterns, such as the ``two degrees
separation''\citep{williams_two_2002} and skewed or power law degree
distributions\citep{dunne_food-web_2002}, are found on binary food
webs. However, these studies never considered weights, which may
hide important information of energy flux transferred between
different species\citep{zhang_scaling_2010}.

Weights on edges and nodes stand for energy-matter flux between two
compartments and biomass on each unit
respectively\citep{higashi_network_1993}. Interestingly,
\cite{lindeman_trophic-dynamic_1942} and
\cite{odum_self-organization_1988}'s seminal works on food webs are
all based on weighted networks. \cite{patten_energy_1985} et al.
further developed a systematic method called environ analysis to
uncover some hidden information in energy flows on networks. Besides
the common phenomena found in earlier literatures including the
hierarchical trophic structure and pyramid of biomass
distribution\citep{odum_system_1983,e.p._odum_fundamentals_2004},
some quantitative and ubiquitous patterns such as dominant indirect
effects\citep{higashi_network_1993}, network
amplification\citep{patten_trophic_1990}, network
homogenization\citep{patten_trophic_1990}, pathway
proliferation\citep{patten_energy_1985,borrett_functional_2007} and
network synergism\citep{patten_energy_1992} are found by environ
analysis in various food webs\citep{fath_review_1999}.

Metabolic theory is one of the greatest progresses in ecology in
recent years which can be incorporated into food web
studies\citep{brown_toward_2004}. Several universal patterns or laws
related to body size are discovered in the last two
decades\citep{brown_scaling_2000-1,west_review_2005}. For example,
the three quarters power law relationship between metabolism and
body mass (Kleiber's law) is one of the most fundamental laws in
metabolic theory\citep{kleiber_body_1932}. Some ecologists also
tried to link these patterns to trophic
structures\citep{nee_relationship_1991,loeuille_evolution_2006,jennings_abundance_2003,damuth_population_1981,allen_global_2002},
including the energetic equivalence
rule\citep{allen_global_2002,loeuille_evolution_2006}; the
trivariate relationship among body mass, trophic level and energy
flows\citep{cohen_ecological_2003}. Nevertheless, some simple but
important relations, say, biomass v.s. throughflow of each species,
throughflow v.s. input flow and output flow are seldom addressed by
these previous studies.

In parallel with these studies in ecology, the complex network
community also started to pay attention to weight information of
networks in recent
years\citep{barrat_architecture_2004,almaas_global_2004,serrano_extracting_2009,tumminello_tool_2005}.
By incorporating the statistical mechanics method and random graph
theory, weighted network analysis also revealed a series of
universal patterns in various weighted networks, e.g. air traffic
networks\citep{barrat_architecture_2004,guimera_worldwide_2005},
metabolism networks\citep{almaas_global_2004}, world trade
web\citep{bhattacharya_international_2007} and stock-sharing
networks of companies\citep{vitali_network_2011}, etc. The new found
common patterns include: long tailed distribution of node intensity
(total weights of each node), the power law relationship between
degree and intensity\citep{barrat_architecture_2004}, the so called
gravity
law\citep{james_e._anderson_gravity_2011,erlander_gravity_1990,krings_urban_2009},
and so forth. Energy flow networks in ecosystem no doubt are also
weighted networks though the weight here has the special meaning,
i.e. the energy flux transferred by different
species\citep{zhang_scaling_2010}. Therefore, it is reasonable to
conjecture that the patterns found in other weighted networks should
be also suitable for the weighted food webs. This paper tries to
apply the approaches developed by complex weighted network studies
to the energy flow networks in ecology.

This paper is organized as follows: in Section \ref{sec.methods}, we
introduce some basic variables including the weights of edges and
nodes. And also, the so called DGBD (Discrete version of a
Generalized Beta
Distribution)\citep{martinez-mekler_universality_2009} curve which
can fit the weights distributions better than the traditional curves
is introduced. After that, the results of biomass and energy flow
distributions and several universal relationships including the
allometric law at the population level, the dissipation law and the
gravity law on 19 weighted food webs are shown in Section
\ref{sec.results}. After that, several interesting problem around
these common patterns are discussed. We found an interesting
negative linear relationship between exponents of fitted DGBD
distributions in energy flow, biomass and degree distributions (see
Section \ref{sec.exponent}). And the mathematical relations among
the scaling exponents are derived. Finally, the connection with the
abundance-body mass relationship is discussed in Section
\ref{sec.discussion}.
\section{Materials and Methods}
\label{sec.methods}
\subsection{Data source}
We have investigated 19 food webs in different ecological
environments. The food web information includes node (species or
non-living compartment), node weight (biomass of a node), edge
(energy flow relationship but not feeding relationship), and edge
weight (the amount of energy flow from node $i$ to $j$). The energy
flow between two nodes was measured as the unit volume flow of the
carbon element into or out of the node (the unit is gC/$m^2$/year)).
The biomass stands for the total mass of living biological organisms
of a species in a certain period of time and per unit volume.
Customarily it was also measured by carbon content (the unit is
gC/$m^2$) \citep{baird_seasonal_1989}. These food webs' information
is obtained from the online database\footnotemark
\footnotetext[1]{http://vlado.fmf.uni-lj.si/pub/networks/data/bio/foodweb/foodweb.htm},
which is based on the published
papers\citep{baird_assessment_1998,baird_seasonal_1989,ulanowicz_growth_1986,almunia_benthic-pelagic_1999,monaco_comparative_1997,hagy_eutrophication_2002}.
In Table \ref{tab.foodwebs}, we list the name and the number of
nodes $N$ and edges $E$ of each web.

\begin{table}
\centering
 \caption{Empirical food webs and their topological
 properties}{\small ($N$ stands for the number of vertices of the network and $E$ is the number of edges. The webs are sorted by $E$.)}
 \label{tab.foodwebs}
\begin{tabular}{llllll}
\hline Food web & Abbre.  & $N$ & $E$\\
\hline

Crystal River Creek (Delta Temp) &CrystalD  & 23 & 60\\
Crystal River Creek (Control) &CrystalC  & 23 & 81\\
Chesapeake Bay Mesohaline Net &Chesapeake  & 38 & 122\\
Lower Chesapeake Bay in Summer & ChesLower & 36 & 115\\
Middle Chesapeake Bay in Summer &ChesMiddle  & 36 & 149\\
Upper Chesapeake Bay in Summer &ChesUpper & 36 & 158\\
Narragansett Bay & Narragan  & 34 & 158\\
Lake Michigan & Michigan  & 38 & 172\\
St. Marks River (Florida) & StMarks  & 53 & 270\\
Mondego Estuary - Zostrea site & Mondego  & 45 & 348\\
Cypress, Wet Season & CypWet  &70 & 545\\
Cypress, Dry Season & CypDry  & 70 & 554\\
Everglades Graminoids, Dry Season  & GramDry  & 68 & 793\\
Everglades Graminoids, Wet Season  & GramWet  & 68 & 793\\
Mangrove Estuary, Dry Season & MangDry  & 96 & 1339\\
Mangrove Estuary, Wet Season & MangWet  & 96 & 1440\\
Florida Bay, Wet Season & BayWet  & 127 & 1938\\
Florida Bay, Dry Season & BayDry  & 127 & 1969\\
Florida Bay & Florida & 127 & 1938\\
 \hline
\end{tabular}
\end{table}
\subsection{Basic variables}
Our work is based on the flux matrix of a weighted food web. An
ecological energy flow network is a weighted directed graph that
represents relationships of ecological energy transfer between
species. This graph can be represented by a flux matrix:
\begin{equation}\label{eq2-1}
F_{(N+2)\times (N+2)}\;=\{f_{ij}\}_{(N+2)\times (N+2)},
\;\;\;\forall i,j\in [0,N+1]
\end{equation}
where $f_{ij}$ is the energy flow from species $i$ to $j$. Two
special nodes representing environment (node $0$ and node $N+1$) are
added to the web. Node $0$ denotes the source of energy flow,
whereas node $N+1$ represents the sink. The dissipative and exported
energy flow to the node $N+1$. Therefore, there are totaly
$(N+2)\times (N+2)$ entries. The flux matrix $f_{(N+2)\times (N+2)}$
can be read from the original weighted food
webs\citep{baird_assessment_1998,baird_seasonal_1989,ulanowicz_growth_1986,almunia_benthic-pelagic_1999,monaco_comparative_1997,hagy_eutrophication_2002}.

\begin{figure}
\centerline {\includegraphics{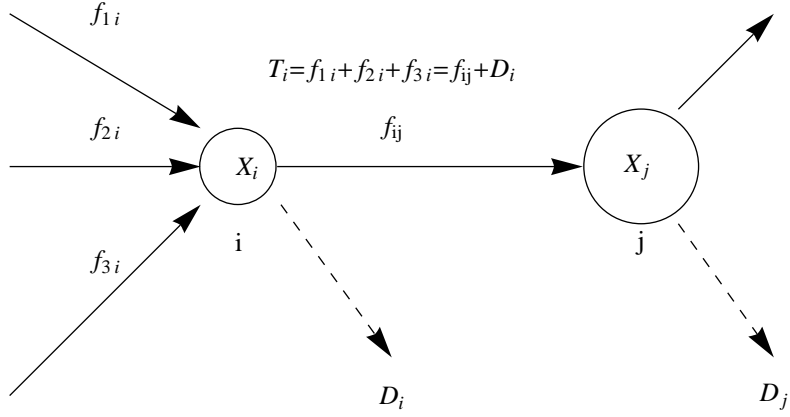}} \vskip3mm \caption{An
Illustration of variables}\label{fig.illustration}
\end{figure}

We can calculate the total through flow of any given node $i$
according to the flux matrix $F_{(N+2)\times (N+2)}$ (see figure
\ref{fig.illustration}). This value is also called node strength in
complex weighted network studies \citep{almaas_global_2004}. Because
the network is always balanced, we need only to calculate the efflux
of each node as $T_i$,
\begin{equation}\label{eq2-2}
T_i\;=\sum_{j=1}^{N+1}f_{ij}=\sum_{j=0}^{N}f_{ji},\;\;\;\forall i\in
[1,N].
\end{equation}

In addition, we define another variable to represent the weight of a
node $X_i$, indicating the biomass of $i$. This information is also
available from the original weighted food webs.

$D_i$ is the dissipated flow to the environment (node $N+1$) from
species $i$: $f_{i,N+1}$ (see figure \ref{fig.illustration}). The
dissipation flow includes the flows of output and respiration.

$k_i^{in}$ and $k_i^{out}$ are in-degree and out-degree, i.e., the
number of inward edges (excluding edges from the source) and outward
edges (excluding the edges to the sink) of $i$ respectively. In the
example network in figure \ref{fig.illustration}, $k_i^{in}=3$ and
$k_i^{out}=1$.

\subsection{Distributions and DGBD curves}
We will study the distributions of $T_i$, $X_i$, $k_i^{in}$ and
$k_i^{out}$ in any empirical food web. Instead of giving the
empirical density or distribution function
\citep{zhang_scaling_2010}, we use the rank-ordered curve to show
the distributions of these variables. For example, if we have a
small food web with 5 species, and their biomass values are $\{100,
19, 200, 5, 1\}$ gC/$m^2$. A sequence of biomass values in a
decreasing order can be obtained: $\{200,100,19,5,1\}$. Then we plot
this sequence on a coordinate with the horizontal axis as the rank
value, namely $\{1,2,3,4,5\}$ and the vertical axis as the biomass
values. So the final curve on this coordinate is the rank-ordered
curve. The main advantage of adopting this curve is its simplicity
for the calculation, as well as it contains the same information as
the distribution function \citep{newman_power_2005}.

Then, we use the DGBD (Discrete version of a Generalized Beta
Distribution) function to fit the rank-ordered curve:
\begin{equation}\label{eq2-3}
Y(r_i)=A\frac{(N+1-r_i)^a}{r_i^b},\hspace{1cm} a,b\geqslant 0
\end{equation}
where, $Y(r_i)$ is the value of the concerned variable ($T_i$ or
$X_i$) of the node $i$, and $r_i$ is the decreasing order of $i$
ranked by $Y(r_i)$ values. $N$ is the total number of nodes in the
food web. $A,a,b\geq 0$ are parameters to be estimated. $A$ stands
for the magnitude of flow or biomass in this food web which is
dependent on the measurement units. $a$ and $b$ are exponents of
power laws in the tail and head part of the curves respectively. If
we set $a=0$, then formula (\ref{eq2-3}) becomes $Y(r_i)=A/{r_i^b}$,
which is the famous Zipf's law \citep{newman_power_2005}. However,
the classic Zipf's law is not always the best choice for fitting
empirical data due to the large deviations in the tail part of the
rank-ordered curves \citep{martinez-mekler_universality_2009}. This
disadvantage can be mended by introducing a new exponent $a$ in the
DGBD fitting. Previous study shows that formula (\ref{eq2-3}) can
fit lots of empirical data very well
\citep{martinez-mekler_universality_2009}.

\begin{figure}
\centerline {\includegraphics[scale=0.95]{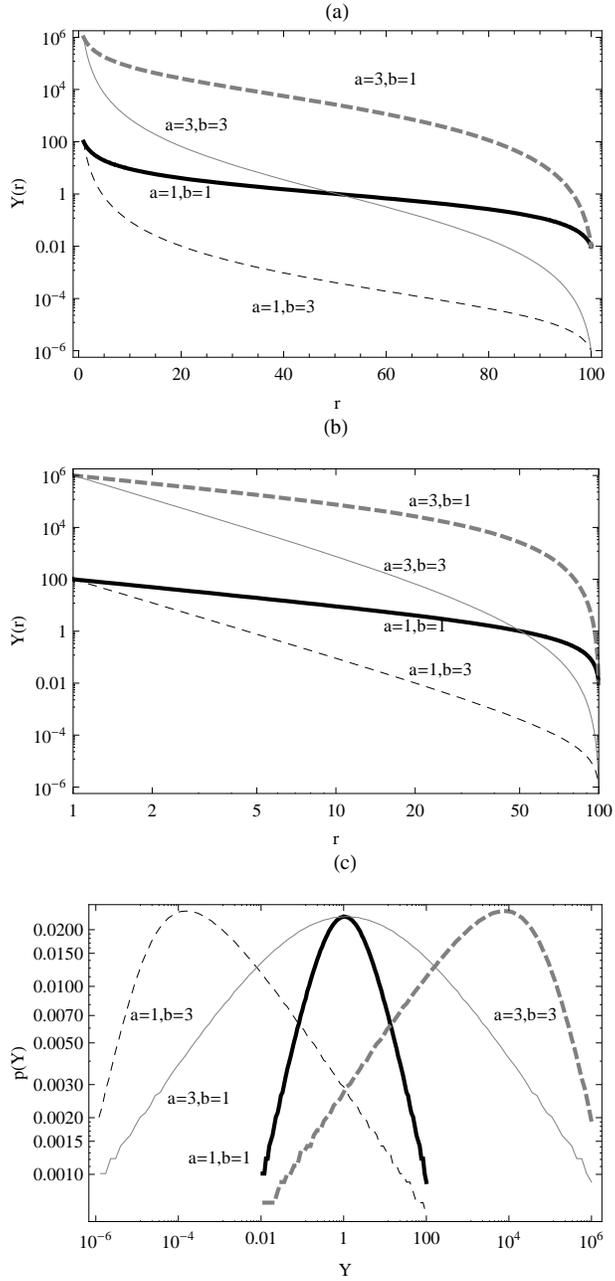}} \vskip3mm
\caption{Different shapes of rank-ordered curves with different $a$
and $b$ under the same $A=1$ and $N=100$. (a) Rank-ordered curves on
a linear-log plot; (b) The same rank-ordered curves on a log-log
plot; (3) The probability density curves of the same four groups of
data on a log-log plot}\label{fig.dgbd}
\end{figure}

Notice that, if we set $a=0$ in equation \ref{eq2-3}, then the
rank-ordered curve of a variable $Y$ follows Zipf's law with the
exponent $b$. Therefore, the density function of the random variable
$Y$ is a power law with the exponent $1/b+1$ due to the one-one
correspondence between the rank-ordered function and probability
density function\citep{newman_power_2005}. So, DGBD curve is capable
to not only fit those data with power law tails (note that the tail
part of the density function is just the head part of the
rank-ordered curve) but also the data with an obvious deviation from
power laws by tuning the extra exponent $a$ (figure \ref{fig.dgbd}).

As shown in figure \ref{fig.dgbd}, different combinations of $a$ and
$b$ correspond different shapes of the distribution curve. Exponent
$b$ is the slope of the head part of the rank-ordered curve in
figure \ref{fig.dgbd} (b) as well as the tail part of the
probability density curve in figure \ref{fig.dgbd} (c). Therefore,
we say $b$ indicates the heterogeneity of $Y$ distributing for the
large species. On the other hand, $a$ indicates the unevenness of
the tail part of the rank-ordered curve (figure \ref{fig.dgbd} (a)
and (b)) as well as the head part of the probability density curve
(figure \ref{fig.dgbd} (c)). As $a$ increases, the tail of the
rank-ordered curve drops very fast so that the $Y$ sharing
heterogeneity in the small species is very large.

We mainly adopt OLS (Ordinary Linear Square regression) method to
fit DGBD curves and other power law relationships. We can take
logarithmic on equation \ref{eq2-3} to derive:

\begin{equation}\label{eq2-3}
\log(Y(r_i))=\log(A)+a \log(N+1-r_i)-b \log(r_i).
\end{equation}

Thus, $\log(Y(r_i))$ depends on two variables $\log(N+1-r_i)$ and
$\log(r_i)$ so that the bivariate linear regression method can be
used to derive the coefficients $\log(A), a$ and $b$.

\section{Results}
\label{sec.results}
\subsection{The DGBD distributions}
\label{sec.dgbd}

\begin{figure}
\centerline {\includegraphics[width=20cm]{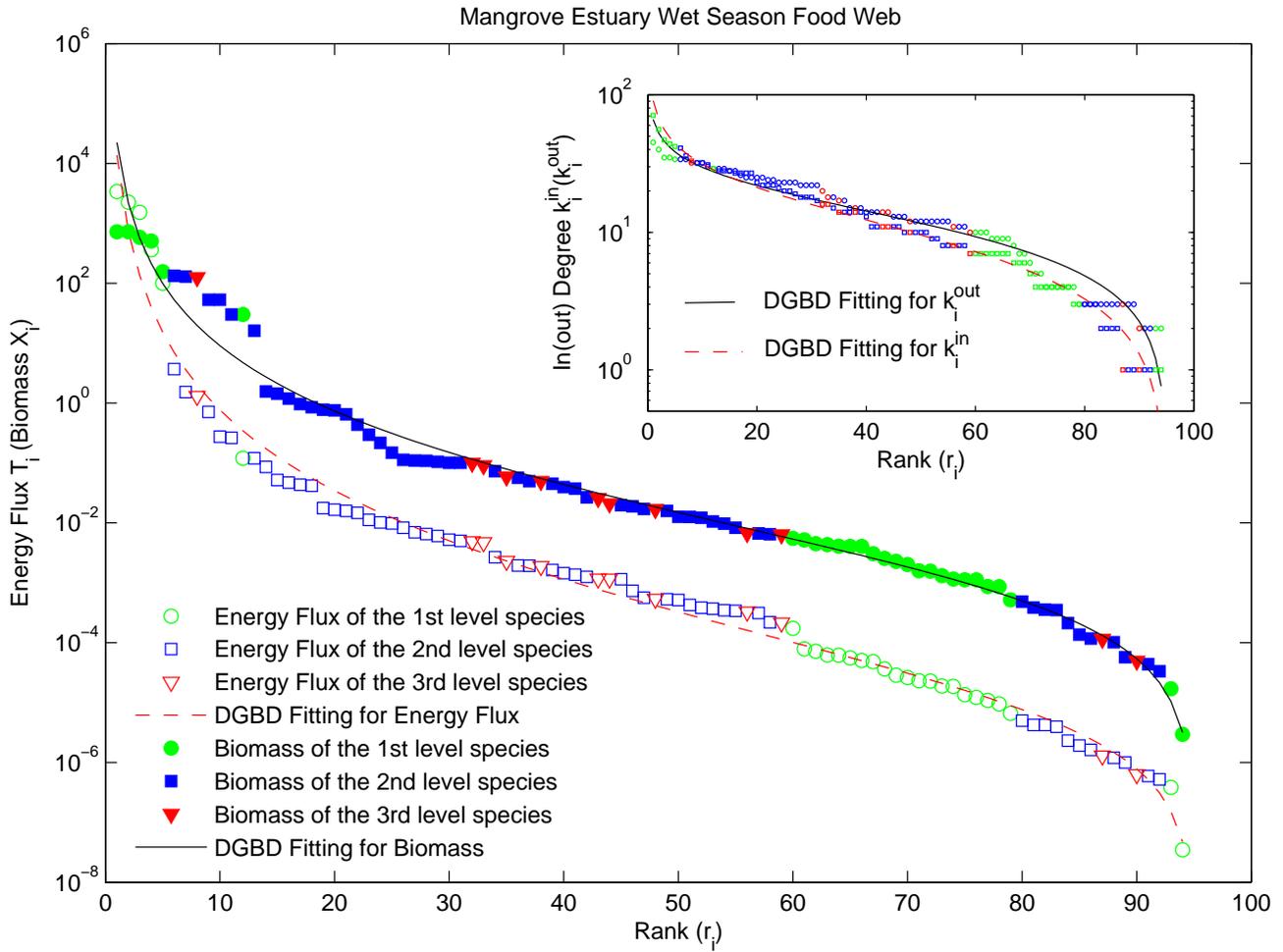}} \vskip3mm
\caption{Rank-ordered distributions and DGBD fittings of $T_i, X_i,
k_i^{out}$ and $k_i^{in}$ for Mangwet food web. The indications of
1st, 2nd and 3th level species in the legend represent the first,
second and third trophic level species in the food
web.}\label{fig.distributions}
\end{figure}

We obtain the distribution of $T_i,X_i,k_i^{in}$ and $k_i^{out}$ for
each of the 19 empirical food webs, and fit them by DGBD curves. One
selected food web as an example is plotted in figure
\ref{fig.distributions}.

From figure \ref{fig.distributions}, we know the curves can be
divided into three parts by two inflexion points, and each part
obeys independent logarithmic decreasing behavior. Obviously, the
head and tail parts of the curve have much steeper slopes than the
middle part. The local slope of the curves indicates the
heterogeneities of the throughflow distribution. In other words, the
larger the absolute value of slope is, the higher the degree of
heterogeneity of a vertex correspondingly is. Therefore, we can
conclude that the nodes in the heads and tails are more
heterogeneous than the ones in the middle.

In figure \ref{fig.distributions}, we distinguish nodes by their
trophic levels. Green circles, blue squares, and red triangles
represent the first, second and third trophic level species
respectively. It is observed that nodes of first trophic level
locate both at the head and tail parts, while most of the second
level species locate at the middle part, and most third trophic
level species are in the middle or tail parts of the curve. This
distribution pattern of species on different trophic levels is
similar for all large food webs (the food webs below Michigan web in
table \ref{tab.foodwebs}). We may conclude that the distribution of
throughflow on the second trophic level is much more even than the
first and third trophic levels. The inset figure of figure
\ref{fig.distributions} plots the rank-ordered distributions of
in-degree and out-degree of all edges. These two curves have the
similar shape with the energy flow and biomass distributions.

\begin{table}
\centering
 \caption{DGBD fitting parameters for different variables}{\small (The webs are sorted by their number of edges.)}
 \label{tab.dgbd}
\begin{tabular}{lllllllllll}
\hline Web & $a_{T_i}$& $b_{T_i}$& $R_{T_i}^2$ & $a_{X_i}$ & $b_{X_i}$ & $R_{X_i}^2$ & $a_{k_i^{out}}$ & $b_{k_i^{out}}$ & $a_{k_i^{in}}$ & $b_{k_i^{in}}$\\
\hline
CrystalD&$0.45$&$3.96$&$0.96$&$0.00$&$4.21$&$0.97$&$0.00$&$0.69$&$0.00$&$0.98$\\
CrystalC&$0.76$&$3.41$&$0.97$&$0.24$&$3.89$&$0.95$&$0.20$&$0.56$&$0.35$&$0.71$\\
Chesapeake&$2.93$&$1.54$&$0.94$&$1.70$&$2.17$&$0.99$&$0.23$&$0.50$&$0.19$&$0.73$\\
ChesLower&$10.02$&$0.00$&$0.80$&$5.53$&$0.75$&$0.85$&$0.68$&$0.17$&$0.15$&$0.88$\\
ChesMiddle&$6.88$&$0.00$&$0.83$&$3.42$&$0.77$&$0.87$&$0.55$&$0.25$&$0.40$&$0.63$\\
ChesUpper&$4.77$&$0.00$&$0.72$&$3.42$&$0.86$&$0.88$&$0.61$&$0.16$&$0.37$&$0.70$\\
Narragan&$0.80$&$2.54$&$0.93$&$0.16$&$1.88$&$0.93$&$0.21$&$0.47$&$0.41$&$0.56$\\
Michigan&$6.59$&$0.81$&$0.95$&$4.80$&$0.18$&$0.89$&$0.51$&$0.47$&$0.45$&$0.45$\\
StMarks&$1.25$&$1.33$&$0.99$&$0.94$&$2.14$&$0.98$&$0.59$&$0.38$&$0.54$&$0.40$\\
Mondego&$3.36$&$1.70$&$0.97$&$1.41$&$2.43$&$0.99$&$0.57$&$0.26$&$0.48$&$0.94$\\
Cypwet&$2.74$&$2.63$&$0.97$&$3.03$&$4.23$&$0.98$&$0.41$&$0.56$&$0.62$&$0.66$\\
Cypdry&$2.34$&$2.36$&$0.95$&$2.51$&$3.99$&$0.98$&$0.43$&$0.52$&$0.63$&$0.66$\\
Gramdry&$2.08$&$3.18$&$0.97$&$2.75$&$2.75$&$0.97$&$0.34$&$0.43$&$0.67$&$0.54$\\
Gramwet&$3.03$&$2.94$&$0.97$&$3.74$&$2.40$&$0.97$&$0.34$&$0.43$&$0.67$&$0.54$\\
Mangdry&$1.51$&$3.22$&$0.97$&$1.41$&$4.08$&$0.98$&$0.66$&$0.32$&$0.79$&$0.42$\\
Mangwet&$1.67$&$3.32$&$0.98$&$1.63$&$4.18$&$0.99$&$0.66$&$0.32$&$0.79$&$0.43$\\
Baywet&$1.65$&$2.80$&$0.97$&$2.26$&$2.40$&$0.99$&$0.74$&$0.40$&$0.97$&$0.25$\\
Baydry&$1.62$&$2.63$&$0.98$&$2.24$&$2.28$&$0.98$&$0.75$&$0.39$&$0.97$&$0.25$\\
Florida&$1.65$&$2.80$&$0.97$&$2.26$&$2.40$&$0.99$&$0.74$&$0.40$&$0.97$&$0.25$\\
\hline
\end{tabular}
\end{table}

In Table \ref{tab.dgbd}, we list all the fitted parameters and
$R^2$s of DGBD for the 19 webs. Only the exponents of $a$ and $b$
are shown in the table because $A$ representing the magnitude of
throughflow or biomass is relatively unimportant. By comparing
different rows, we know that the food webs with more edges can be
better fitted by DGBD curves because their $R^2$s are larger. Notice
that there are three food webs (ChesLower, ChesMiddle and ChesUp)
having exponents $b_{X_i}=0$ and one web (CrystalD) having exponents
$a_{X_i},a_{k_i^{in}},a_{k_i^{in}}=0$. That means these webs are
anomalous and reach the extremal cases of DGBD curves. However the
$R^2$'s of the the fitting curves are also very high.

The last four columns show the DGBD fittings of in and out degrees.
An obvious trend is the exponent $a_{k_i}$ increases, however
$b_{k_i}$ decreases with the size of the food web. That indicates
the number of connected species distributes on different nodes more
evenly in small food webs and the average degrees increase with the
size of the web.

\subsection{Allometric scaling laws at the population level}
\label{sec.powerlaws}

\begin{figure}
\centerline {\includegraphics[scale=0.9]{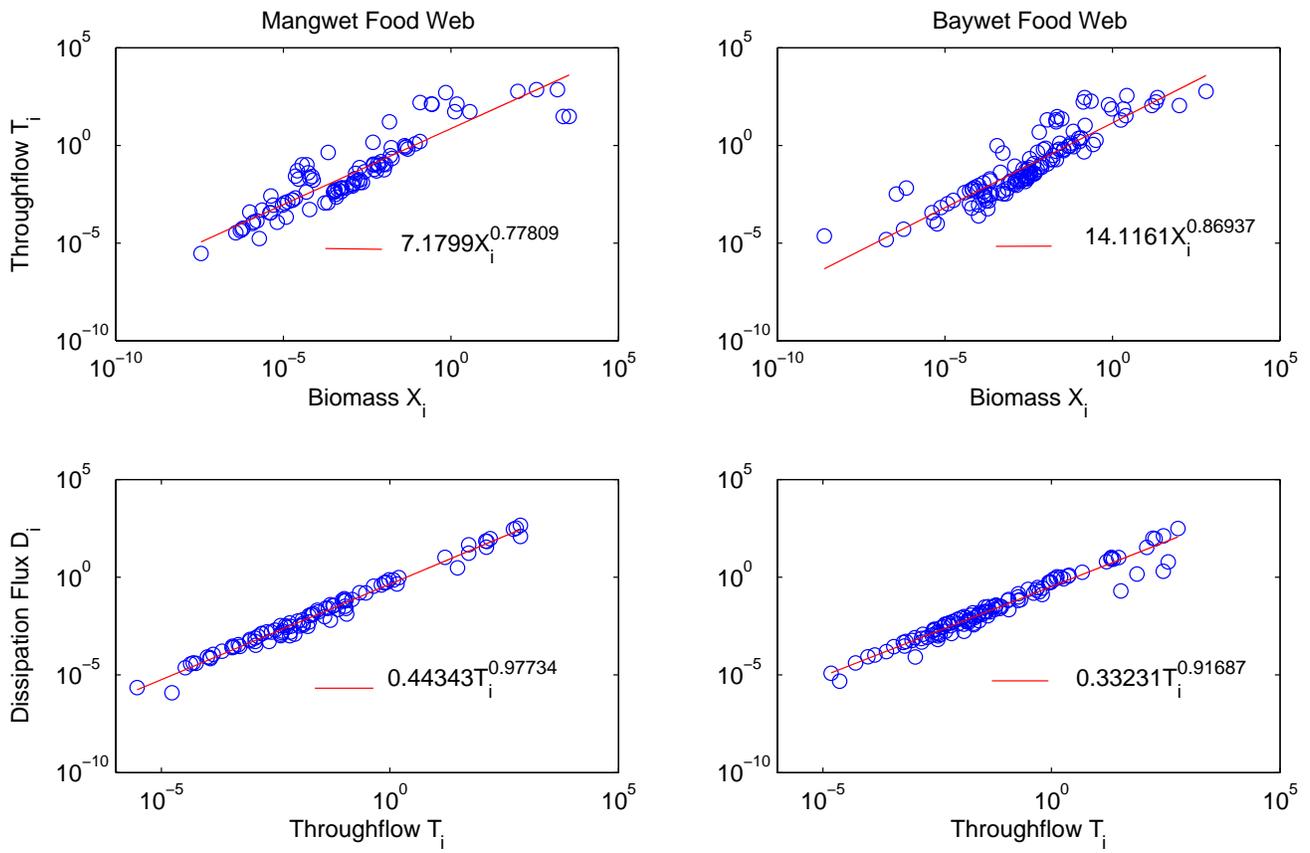}} \vskip3mm
\caption{Power law relationships of $T_i$ v.s. $X_i$ and $D_i$ v.s.
$X_i$. The original data points as well as the OLS fittings are
shown on the log-log coordinate}\label{fig.powerlaws}
\end{figure}

According to the similarity between the distribution curves of $T_i$
and $X_i$ in figure \ref{fig.distributions}, we guess there may
exist a connection between variables $X_i$ and $T_i$. The log-log
plot of these two variables shows that the relationship between
$X_i$ and $T_i$ is actually a power law:
\begin{equation}\label{eq2-4}
T_i\propto X_i^{\alpha},
\end{equation}
where exponent $\alpha$ is a parameter to be estimated for each food
web. As shown in upper row in figure \ref{fig.powerlaws}, the sample
points aggregate around their fitted lines very well. This
relationship is ubiquitous for all 19 food webs as shown in Table
\ref{tab.powerlaws}. We use the ordinary linear regression method to
find the best fitting lines (Table \ref{tab.powerlaws}). And most
$R^2$'s are larger than $0.8$. The $R^2$'s and exponents decrease
with the scale of the network because the statistical significance
increases with the number of samples.

\begin{table} \centering
 \caption{Fitting exponents and goodness of power law relationships }{\small ( The webs are sorted by their number of edges.)}
 \label{tab.powerlaws}
\begin{tabular}{llllllllllll}
\hline Web & $\alpha$& $R_{allo}^2$ &$c$ & $\beta$ & $R_{diss}^2$ & $\gamma$ &$R_{gra}^2$ & $\gamma_1$ & $\gamma_2$ & $R_{bi}^2$&$\eta$\\
\hline
CrystalD&$0.98$&$0.90$&$0.67$&$0.96$&$1.00$&$0.63$&$0.70$&$0.57$&$0.75$&$0.74$&11.5\\
CrystalC&$0.92$&$0.88$&$0.68$&$0.96$&$0.99$&$0.53$&$0.65$&$0.50$&$0.57$&$0.65$&2.125\\
Chesapeake&$1.03$&$0.75$&$0.50$&$0.99$&$0.98$&$0.68$&$0.84$&$0.62$&$0.77$&$0.85$&-9.33\\
ChesLower&$1.75$&$0.91$&$0.49$&$0.95$&$0.99$&$0.70$&$0.75$&$0.61$&$0.84$&$0.76$&-1.33\\
ChesMiddle&$1.58$&$0.82$&$0.47$&$0.88$&$0.85$&$0.67$&$0.77$&$0.60$&$0.78$&$0.78$&-1.43\\
ChesUpper&$1.21$&$0.64$&$0.58$&$0.95$&$0.99$&$0.64$&$0.64$&$0.62$&$0.67$&$0.64$&-2.19\\
Narragan&$1.17$&$0.55$&$2.05$&$0.81$&$0.94$&$0.54$&$0.81$&$0.49$&$0.60$&$0.81$&-2.47\\
Michigan&$1.20$&$0.94$&$0.67$&$0.99$&$1.00$&$0.62$&$0.86$&$0.57$&$0.72$&$0.87$&-2.25\\
StMarks&$0.69$&$0.70$&$0.43$&$0.99$&$0.95$&$0.68$&$0.74$&$0.76$&$0.56$&$0.75$&-0.19\\
Mondego&$1.24$&$0.87$&$0.53$&$0.98$&$1.00$&$0.79$&$0.85$&$0.83$&$0.70$&$0.86$&-2.04\\
Cypwet&$0.66$&$0.78$&$0.46$&$0.97$&$0.99$&$0.70$&$0.84$&$0.85$&$0.55$&$0.87$&-0.26\\
Cypdry&$0.63$&$0.76$&$0.41$&$0.96$&$0.95$&$0.68$&$0.81$&$0.81$&$0.57$&$0.83$&-0.32\\
Gramdry&$0.90$&$0.90$&$0.58$&$0.97$&$1.00$&$0.66$&$0.76$&$0.61$&$0.73$&$0.77$&1.5\\
Gramwet&$0.93$&$0.92$&$0.59$&$0.98$&$1.00$&$0.71$&$0.81$&$0.66$&$0.79$&$0.81$&2.57\\
Mangdry&$0.77$&$0.82$&$0.45$&$0.98$&$0.98$&$0.58$&$0.77$&$0.60$&$0.56$&$0.77$&0.09\\
Mangwet&$0.78$&$0.83$&$0.44$&$0.98$&$0.98$&$0.59$&$0.77$&$0.60$&$0.57$&$0.77$&0.13\\
Baywet&$0.87$&$0.81$&$0.33$&$0.92$&$0.95$&$0.62$&$0.79$&$0.67$&$0.54$&$0.80$&0.92\\
Baydry&$0.85$&$0.81$&$0.32$&$0.91$&$0.95$&$0.61$&$0.78$&$0.68$&$0.52$&$0.78$&0.67\\
Florida&$0.87$&$0.81$&$0.33$&$0.92$&$0.95$&$0.62$&$0.79$&$0.67$&$0.54$&$0.80$&0.92\\
\hline
\end{tabular}
\end{table}

This specific scaling relationship reminds us the famous allometric
scaling law \citep{brown_toward_2004, west_review_2005,
banavar_size_1999}. \cite{kleiber_body_1932} found that the
metabolism and body size of an organism usually follows a ubiquitous
power law relationship with an exponent around $3/4$. If we treat
the whole population of a species as an integrated organism, then
$T_i$ is its metabolism and $X_i$ is its body mass. Therefore,
equation \ref{eq2-4} can be viewed as the allometric scaling law at
the population level. Nevertheless, unlike the universal Kleiber's
law for species, the allometric scaling exponents of population on
food webs are not universal but fluctuate in between $[0.63, 1.75]$.

\subsection{Dissipation law}
\label{sec.dissipationlaw}

As pointed by the earlier ecological
studies\citep{odum_system_1983,lindeman_trophic-dynamic_1942}, a
large fraction of energy flows dissipates to the environment in the
whole ecosystem. The dissipated energy flow can be captured by the
variable $D_i$ which is also available from the original data.
Empirically, $D_i$ scales with throughflow $T_i$ in the following
way:
\begin{equation}\label{eq.dissipation}
D_i=c T_i^{\beta}.
\end{equation}
Equation \ref{eq.dissipation} is called dissipation law in this
paper, where $c$ and $\beta$ are parameters to be estimated. We
observe that the estimated exponents $\beta$ are all slightly
smaller than 1 (see the 2nd row of figure \ref{fig.powerlaws} and
the 4th~6th columns of table \ref{tab.powerlaws}), therefore the
dissipation rate (dissipation per throughflow) decreases with the
throughflow of the species slightly. If $\beta=1$, $c$ is the
average energy dissipation rate of the whole food web. Since the
empirical exponents in table \ref{tab.powerlaws} are approaching to
1, the coefficient $c$'s are almost the dissipation rate of the
specific food web. From table \ref{tab.powerlaws}, we can read $c$'s
are fractional numbers that are smaller than one except the food web
Narragan whose exponent $\beta$ deviates 1 significantly.

\subsection{Gravity Law}
\label{sec.gravitylaw}

Although we have studied the ways of energy flows correlate with the
biomass and dissipation, we still don't know how the energy flows
distribute among different pairs of species. Actually, the large
throughflow nodes can exchange large energy flows. This effect is
reflected by the so called gravity law, namely, the energy flow
between $i$ and $j$ scales with the product of the total
throughflows of $i$ and $j$, i.e.,

\begin{equation}\label{eq.gravity}
f_{ij}\propto (T_iT_j)^{\gamma}.
\end{equation}
This scaling relation is known as the gravity law in other complex
systems. Researchers found the flows, say traffic flow or trade flow
between cities or countries scales with $(m_1 m_2)^{\gamma}/d^\tau$,
where $m_i$ is the size of the system (population of a city or GDP
of a country) and $d$ is the distance between the two systems,
$\gamma$ and $\tau$ are fitting exponents. In our case, the
throughflow of each node $T_i$ is treated as the size of a node
comparable to the population of a city. However, the distance $d$
has no correspondence because spatial information is not included in
our food webs.

\begin{figure}
\centerline {\includegraphics[scale=0.9]{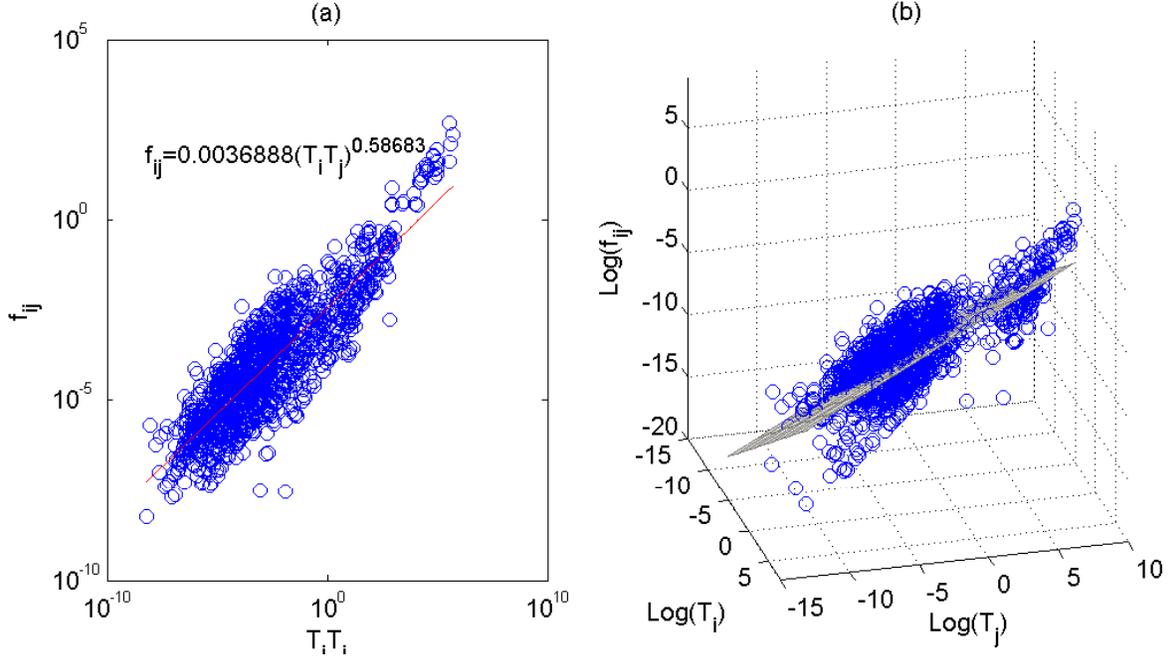}} \vskip3mm
\caption{Univariate and bivariate gravity law of Mangwet food web.
(a) The univariate scaling relationship $f_{ij}\propto
(T_iT_j)^{0.587}$; (b) The bivariate scaling relationship
$f_{ij}\propto T_i^{0.6}T_j^{0.57}$. The original data points as
well as the OLS fittings are shown on the log-log
coordinate}\label{fig.gravitylaws}
\end{figure}

Figure \ref{fig.gravitylaws}(a) shows this phenomenon and the
parameters are listed in the last two columns of table
\ref{tab.powerlaws}. This pattern is not as significant as the
previous two patterns because the $R^2$'s are always smaller than
$0.85$. From figure \ref{fig.gravitylaws} (a), we could observe that
there are several straight bands in the clusters of data points
which indicates that the energy flow between two nodes $f_{ij}$ may
be predicted by other variables rather than the product of $T_i$ and
$T_j$.

As the studies in gravity laws, we suggest that the following
bivariate scaling relation holds,

\begin{equation}\label{eq.bigravity}
f_{ij}\propto T_i^{\gamma_1}T_j^{\gamma_2}.
\end{equation}

Where, $\gamma_1$ and $\gamma_2$ are two estimated parameters.
Figure \ref{fig.gravitylaws} (b) shows this bivariate gravity law.
All the estimated exponents $\gamma_1$, $\gamma_2$ and the
corresponding $R^2$ are shown in the last 2nd, 3rd and 4th columns
in table \ref{tab.powerlaws}. The bivariate gravity law equation
\ref{eq.bigravity} can fit the original data better than equation
\ref{eq.gravity} since the corresponding $R_{bi}^2$'s are slight
higher than $R_{gra}^2$'s. Another interesting phenomenon observed
from table \ref{tab.powerlaws} is that the exponent $\gamma_2$ is
larger than $\gamma_1$ for small food webs but smaller than it for
large food webs except Gramdry and Gramwet. We can conclude that the
large energy flows prefer to link nodes with large throughflow in
all food webs. The dependence of energy flow on the donator (prey)
and receptor (predator) is asymmetric. As the energy flow along each
edge increases, the energy throughflows of receptors increases
faster than donators in small food webs. The speed of increasing of
throughflows for donator is higher than receptors in large food
webs.

All of these observed patterns of energy flows exhibit statistical
significance and universality for all 19 empirical food webs.

\section{Discussions}
\label{sec.discussion}
\subsection{Relationships of exponents}
\label{sec.exponent}

\begin{figure}
\centerline {\includegraphics{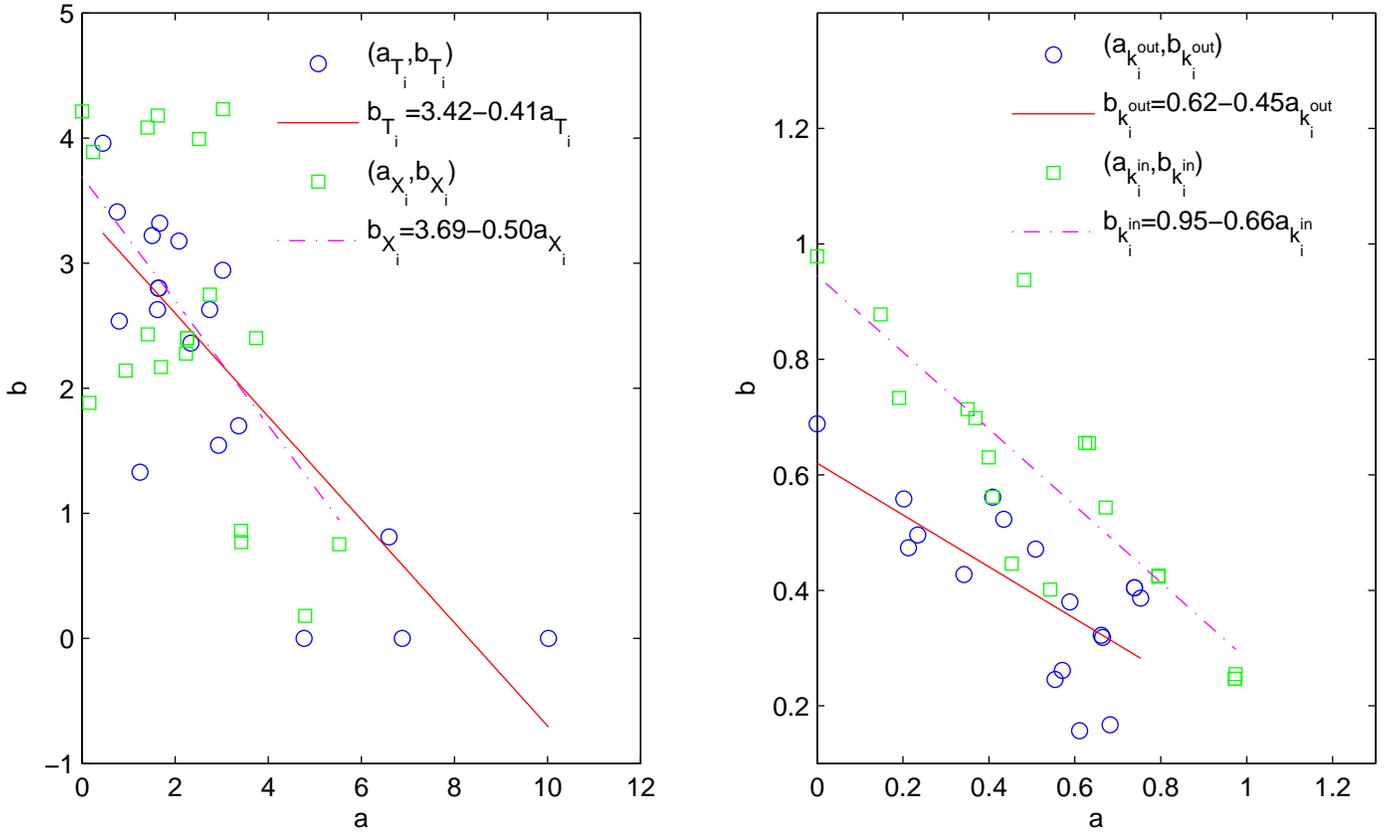}} \vskip3mm \caption{The
linear relationships of DGBD fitted exponents $a$ and $b$ in four
distributions ($T_i,X_i,k_i^{out},k_i^{in}$). The $R^2$'s of the
best fitting lines are
$R_{T_i}^2=0.51,R_{X_i}^2=0.72,R_{k_i^{out}}^2=0.67,R_{k_i^{in}}^2=0.35$
} \label{fig.exponents}
\end{figure}

In section \ref{sec.dgbd} and table \ref{tab.dgbd}, we have obtained
a set of fitting exponents of DGBD curves. These exponents also have
some patterns. It is observed from table \ref{tab.dgbd} that
exponents $a$ and $b$ have a negative correlation. The relationship
is very clear once we plot the pairs of $(a,b)$ in one coordinate
(see figure \ref{fig.exponents}). We separate the $a,b$ exponents of
$T_i,X_i$ and the ones of $k_i^{out}, k_i^{in}$ because the former
has a wider range. However all these pairs of $(a,b)$ show the
nearly linear relationship with similar negative slopes (the mean
slope is $-0.5$) and statistical significance (the $R^2$'s of these
relationships are all larger than $0.5$ except $k_i^{in}$).

We suppose this pattern reflects a kind of particular regularity of
food webs since the similar phenomenon is never reported in previous
studies\citep{martinez-mekler_universality_2009}. The negative
correlation between exponents $a$ and $b$ implies a kind of
complementarity between the heterogeneities of energy flow or
biomass resources distributions at the head and tail part species on
rank curves. As the heterogeneity of the small species increases,
the unevenness of energy distribution in large species decreases.

\begin{figure}
\centerline {\includegraphics{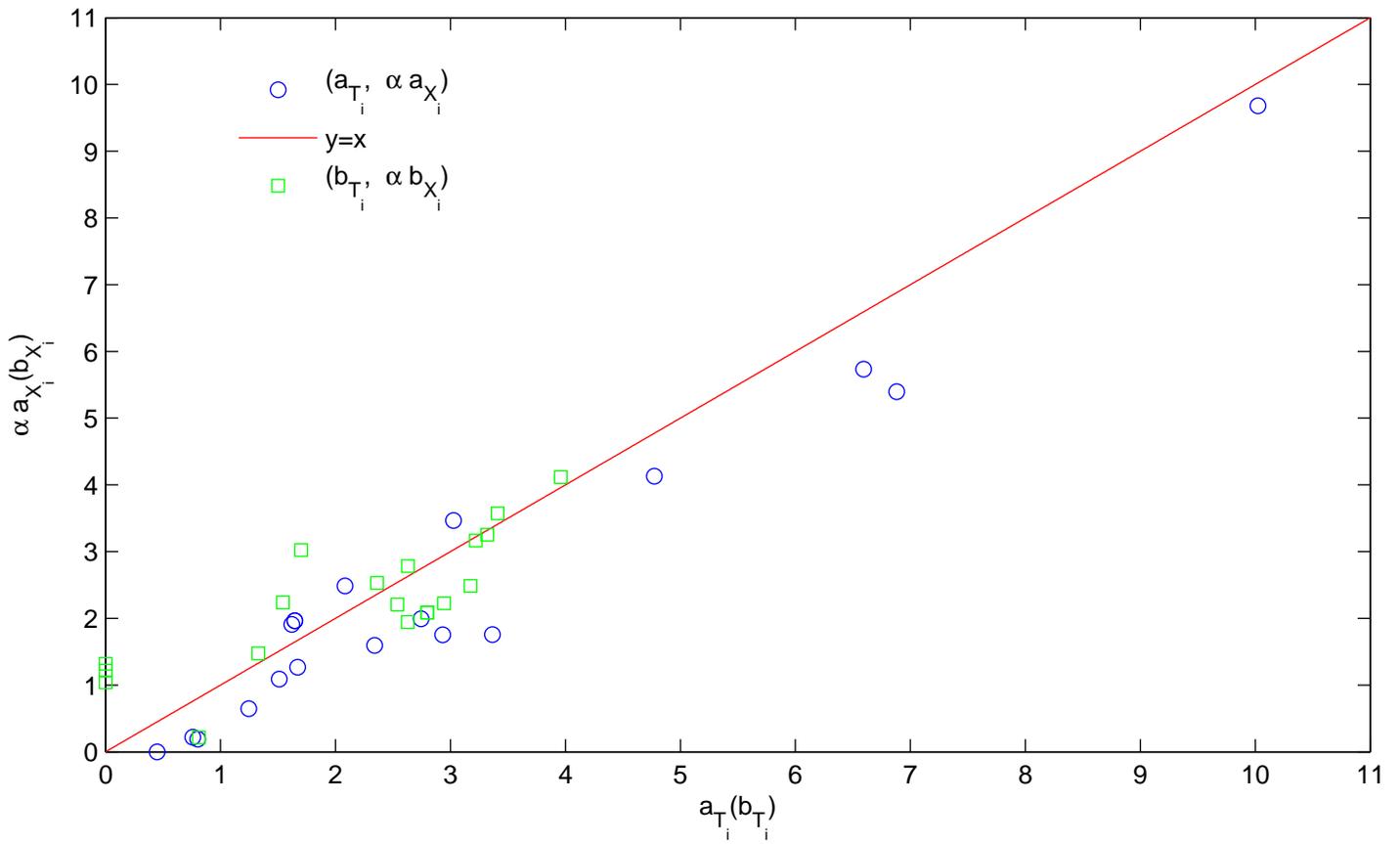}} \vskip3mm \caption{The
real relationship between $a_{T_i}(b_{T_i})$ and
$a_{X_i}\alpha(b_{X_i}\alpha)$, and the predicted one (the 45 degree
diagonal straight line). } \label{fig.exponents2}
\end{figure}

Besides this pattern, other two interesting relationships between
the exponents of distributions and power law relationships can be
derived . We can write down the DGBD distributions of $T_i$ and
$X_i$ in the following forms:
\begin{equation}\label{eq.fidgbd}
T_i\;=A_{T_i}\frac{(N+1-r_i)^{a_{T_i}}}{r_i^{b_{T_i}}}.
\end{equation}
and
\begin{equation}\label{eq.bidgbd}
X_i\;=A_{X_i}\frac{(N+1-r_i)^{a_{X_i}}}{r_i^{b_{X_i}}}.
\end{equation}
for each node $i$. As we have shown in section \ref{sec.powerlaws},
$T_i$ and $X_i$ follow a power law relationship: equation
(\ref{eq2-4}). If we insert equation (\ref{eq.fidgbd}) and
(\ref{eq.bidgbd}) into equation (\ref{eq2-4}), we can easily derive
the following equation:
\begin{equation}\label{eq2-7}
\frac{(N+1-r_i)^{a_{T_i}}}{r_i^{b_{T_i}}}\propto
(\frac{(N+1-r_i)^{a_{X_i}}}{r_i^{b_{X_i}}})^{\alpha}.
\end{equation}
This should be satisfied for any $r_i$. So comparing the
coefficients of the terms $(N+1-r_i)$ and $r_i$, we can derive the
following relationships.
\begin{equation}\label{eq2-9}
a_{T_i}={a_{X_i}}{\alpha},
\end{equation}
and
\begin{equation}\label{eq2-10}
b_{T_i}={b_{X_i}}{\alpha}.
\end{equation}
If these two relations hold for all food webs, the pairs of
$(a_{T_i},\alpha a_{X_i})$ or $(b_{T_i},\alpha b_{X_i})$ in the
empirical food webs should form a straight line with a 45 degree
slope.

From Figure \ref{fig.exponents2}, the pairs of $(a_{T_i},\alpha
a_{X_i})$ or $(b_{T_i},\alpha b_{X_i})$ concentrate around the
predicted relationship with small deviations. That means the
predicted relationships (equations \ref{eq2-9} and \ref{eq2-10}) are
almost correct for the empirical food webs. However, a systemic
deviation from the predicted line exists for $(a_{T_i},\alpha
a_{X_i})$ since all the data points are lower than the theoretical
line. Therefore, we may underestimate the values of $a_{X_i}$ or
overestimate $a_{T_i}$. The errors can not come from the estimation
of $\alpha$ because the similar systemic deviation of $\alpha
b_{X_i}$ is not observed. However the reasons for these errors are
still mysteries for us.

\subsection{Other possible patterns}
Besides the common patterns shown in the previous texts, we have
also investigated other possible patterns exhaustively. However they
are either unclear or trivial.

For example, there is a power law relationship between node degree
and strength in other weighted complex networks as shown in previous
studies \citep{barrat_architecture_2004}. Although a positive
correlation between the degree and strength can be observed in our
food webs, this power law relation is not significant($R^2=0.22$).

The energy flow distributions for each node also obey DGBD curves,
but this content is abandoned in the main text because it can not
provide us more insights.

Another trivial scaling relationship is between the energy flow
$f_{ij}$ and the product of biomass of the two species $X_jX_j$
because it is an obvious result from the scaling relationship in
equation \ref{eq.gravity} and equation \ref{eq2-4}.

Finally, one may guess that a power law relationship between
$k_{in}$ and $k_{out}$ must exist because of the similarity of the
distribution curves in the inset of figure \ref{fig.dgbd}. However,
their relations are not significant once we draw the pairs of
$k_{in}, k_{out}$ in one coordinate. Therefore, all the patterns we
have selected are significant and nontrivial.

\subsection{Abundance and Body Size}
Body size is treated as a very fundamental observable in ecology
because it determines other important variables of organisms
including the trophic
structure\citep{cohen_ecological_2003,brown_toward_2004,brown_ecological_2003}.
\cite{cohen_ecological_2003}, \cite{brown_ecological_2003} discussed
the scaling relationship between abundance and body mass in food
web. We may derive a relationship between abundance and body size
from the scaling relation between throughflow and biomass of each
node. Suppose the numeric abundance of species $i$ is $N_i$, the
average body mass of $i$ is $M_i$, and its average energy metabolism
is $F_i$. Then, according to the definitions,

\begin{equation}\label{eq3-1}
X_i=N_i M_i,
\end{equation}
and
\begin{equation}\label{eq3-2}
T_i=N_i F_i.
\end{equation}

The Kleiber's law links $M_i$ and $F_i$ as follows for any species
$i$,

\begin{equation}\label{eq3-3}
F_i\propto M_i^{3/4}.
\end{equation}

So, insert these relations into equation \ref{eq2-4}, we have,

\begin{equation}\label{eq3-4}
N_i\propto M_i^{\frac{3/4-\alpha}{\alpha-1}}.
\end{equation}

Therefore, the scaling exponent between abundance and body size
$\eta$ is:
\begin{equation}\label{eq3-5}
\eta=\frac{3/4-\alpha}{\alpha-1}.
\end{equation}
This exponent can be estimated according to the exponent $\alpha$
for any empirical food webs. The $\eta$ values are derived in the
last column of Table \ref{tab.powerlaws}. We can see that the
exponent $\alpha$ may be either positive or negative which means the
abundance may increase or decrease with body mass. This conclusion
contradicts with our observations and the exponent deviates from the
previous studies
\citep{cohen_ecological_2003,brown_ecological_2003}. We guess the
problem is the nodes in our webs do not always stand for living
species but other non-living compartments, therefore equation
\ref{eq3-4} may not hold for all nodes in the whole network. As a
result, the abundance-body size scaling exponent can not be
determined by $\alpha$ solely. More discussions on linking trophic
structure, energetics and metabolic theory are deserved for the
future studies.

\section{Concluding Remarks}
The weighted food webs have several common patterns that have never
been found in previous binary food web studies. First, the energy
flow, biomass and degree resources distribute on different species
far from evenness. This heterogeneity can be characterized by a
common rank-ordered curve called DGBD. Second, there are a set of
scaling relationships in weighted food webs. The power law
relationship between $T_i$ and $X_i$ can be regarded as a
counterpart of Kleiber's law in population level. The scaling
relation between the dissipation and throughflow characterize the
dissipative nature of energy flows in ecosystems. And another
interesting common pattern is the so called ``gravity law'' which is
also discovered in other complex systems. Finally, we find two
interesting regularities in the fitted exponents including the
negative correlation between $a$ and $b$ and the predicted
relationship between the distribution exponents and the power law
relation exponent.

This paper only exhibits these common patterns in the empirical
weighted food webs, however the underlying mechanisms that can
reproduce these patterns are left for the future works.



{\bf Acknowledgements}

Thanks for the support of National Natural Science Foundation of
China (No. 61004107). We acknowledge the Pajek web site to provide
food web data online.

 {\bf References}

\bibliographystyle{elsarticle-harv}
\bibliography{ecology}

\end{document}